\documentclass [prl, superscriptaddress, showpacs, twocolumn ]{revtex4}

\usepackage{natbib}
\usepackage{graphicx}
\usepackage{dcolumn}
\usepackage{color}

\newcommand{\grad}{^\circ}
\newcommand{\fevo}{FeVO$_4$}

\newcommand{\fe}{Fe$^{3+}$}

\newcommand{\tna}{T$_{N1}$}
\newcommand{\tnb}{T$_{N2}$}

\begin{document}

\title{Multiferroicity and spiral magnetism in FeVO$_4$ with quenched Fe orbital moments}

\author{A. Daoud-Aladine}
\affiliation{ISIS facility, Rutherford Appleton Laboratory-CCLRC, Chilton, Didcot, Oxfordshire, OX11 0QX, United Kingdom. }
\author{B. Kundys}
\affiliation{Laboratoire CRISMAT-UMR, 6508 ENSI CAEN, 6, Marechal Juin, 14050 Caen, France}
\author{C. Martin}
\affiliation{Laboratoire CRISMAT-UMR, 6508 ENSI CAEN, 6, Marechal Juin, 14050 Caen, France}
\author{P.G. Radaelli}
\affiliation{ISIS facility, Rutherford Appleton Laboratory-CCLRC, Chilton, Didcot, Oxfordshire, OX11 0QX, United Kingdom. }
\affiliation{Dept. of Physics and Astronomy, University College London, Gower Street, London WC1E 6BT, United Kingdom}
\author{P.J. Brown}
\affiliation{Institut Laue Langevin, 6 rue Jules Horowitz, BP 156, 38042 Grenoble Cedex 9, France}
\author{C. Simon}
\affiliation{Laboratoire CRISMAT-UMR, 6508 ENSI CAEN, 6, Marechal Juin, 14050 Caen, France}
\author{L. C. Chapon}
\affiliation{ISIS facility, Rutherford Appleton Laboratory-CCLRC, Chilton, Didcot, Oxfordshire, OX11 0QX, United Kingdom. }

\date{\today}

\begin{abstract}
FeVO$_4$ has been studied by heat capacity, magnetic susceptibility, electric polarization and single crystal neutron diffraction experiments. The triclinic crystal structure is made of \emph{S}-shaped clusters of six Fe$^{3+}$ ions, linked by VO$_4^{3-}$ groups. Two long-range magnetic ordering transitions occur at T$_{N1}$=22K and T$_{N2}$=15K. Both magnetic structures are incommensurate. That stable below T$_{N1}$ is collinear with  amplitude modulated moments whereas below T$_{N2}$ the arrangement is non-collinear with a helicoidal modulation. Below T$_{N2}$, \fevo\ becomes weakly ferroelectric coincidentally with the loss of the collinearity of the magnetic structure. We conclude that \fevo \ provides another example of frustrated spiral magnet similar to the classical TbMnO$_3$ compound. However, \fevo \ has quenched orbital moments and a particular structure clarifying the respective role of anisotropy and magnetic frustration in this type of multiferroic materials.
\end{abstract}


\maketitle \indent There has been a recent surge of interest in a novel class of multiferroic materials\cite{Eerenstein2006}, in which ferroelectricity arises below a magnetic phase transition and as a direct consequence of complex magnetic ordering in systems with strong magneto-electric interactions\cite{Cheong2007}.  This phenomenon, which requires lowering of the magneto-crystalline symmetry to a polar group, can only appear in complex magnetic structures, typically stabilized by magnetic frustration or strong exchange competition, as in Ni$_3$V$_2$O$_8$\cite{Rogado2002} and TbMnO$_3$\cite{Kimura2003}. By contrast, these "novel" or "type II"  multiferroics are simple binary \cite{Kimura2008} or ternary oxides --- a desirable feature for prospective applications. \\
\indent In addition to fulfilling strict magneto-structural symmetry requirements, type-II multiferroics must possess a microscopic mechanism to generate electric dipole moments.  A variety of such mechanisms have been proposed \cite{Jia2007}: magneto-striction is the only allowed mechanism for acentric collinear structures, and is active in the Ising system Ca$_3$(Co,Mn)$_2$O$_6$ \cite{Choi2008} and, most likely, in the commensurate phase of YMn$_{2}$O$_5$\cite{Chapon2006}.  Most other "type-II" multiferroics are cycloidal magnets, where non-collinear spins are key ingredients in the context of the so-called spin-current model \cite{Katsura2005,Mostovoy2006}.  For this, a crucial role is played by relativistic spin-orbit interaction, which can take place at the ligand ionic site, as for pure $e_g$ systems, within the transition-metal $t_{2g}$ orbitals or between $t_{2g}$ and $e_g$ orbitals \cite{Jia2007}.  A particularly interesting case of the latter is provided by high-spin $d^5$ systems (S=5/2) where L=0 in the free ion and the orbital angular momentum is supposedly absent.  As for the acentric magnetic structure itself, this typically comes about when a connected network of superexchange (SE) interactions is destabilized by the presence of either strong next-nearest neighbor interactions or geometrical frustration \cite{Fiebig2002,Eerenstein2006}.\\
\indent In this letter, we describe a new multiferroic compound --- FeVO$_4$ --- in which the magnetic ion is orbitally quenched Fe$^{3+}$ ($d^5$, L=0, S=5/2).  The magnetic and dielectric phase diagram of FeVO$_4$, as determined from magnetization, specific heat and neutron diffraction measurements, is that of a typical cycloidal magnet: ferroelectricity appears below \tnb=15K, coinciding with the appearance of a non-collinear incommensurate magnetic structure (phase II), whereas a second collinear incommensurate magnetic phase (I), stable between \tna$=22K$ and \tnb, is not ferroelectric. Uniquely, FeVO$_4$ does not contain connected magnetic direct exchange or SE paths, and the magnetic modulation is primarily determined by a network of super-super-exchange (SSE) interactions. These pathways form loops connecting an odd number of \fe, suggesting that  frustration plays a key role in promoting non-collinearity and ferroelectricity.\\
\indent Polycrystalline samples and single crystals were prepared following the procedures described in  Refs \onlinecite{Levinson1971} and \onlinecite{Robertson1972}. Electric measurements and specific heat and magnetic susceptibility measurements were carried out on dense pellets of polycrystalline FeVO$_4$ using a PPMS Quantum Design cryostat. The electrical polarization was derived from integration of the pyroelectric current measured with a Keithley 6517 electrometer. Magnetization measurements on single crystals were made with a vertical-field SQUID magnetometer using a horizontal axis sample-rotator. Single crystal neutron nuclear and magnetic Bragg peak intensities were collected  and at T=2K and T=18K on the four-circle diffractometer D15 ($\lambda=1.174\AA$) at the Institut Laue-Langevin (ILL), France. All the nuclear and magnetic structure refinements were carried out with the program FullProF\cite{Rodriguez-Carvajal1993}. \\
\begin{figure}[t]
\includegraphics[width=240pt]{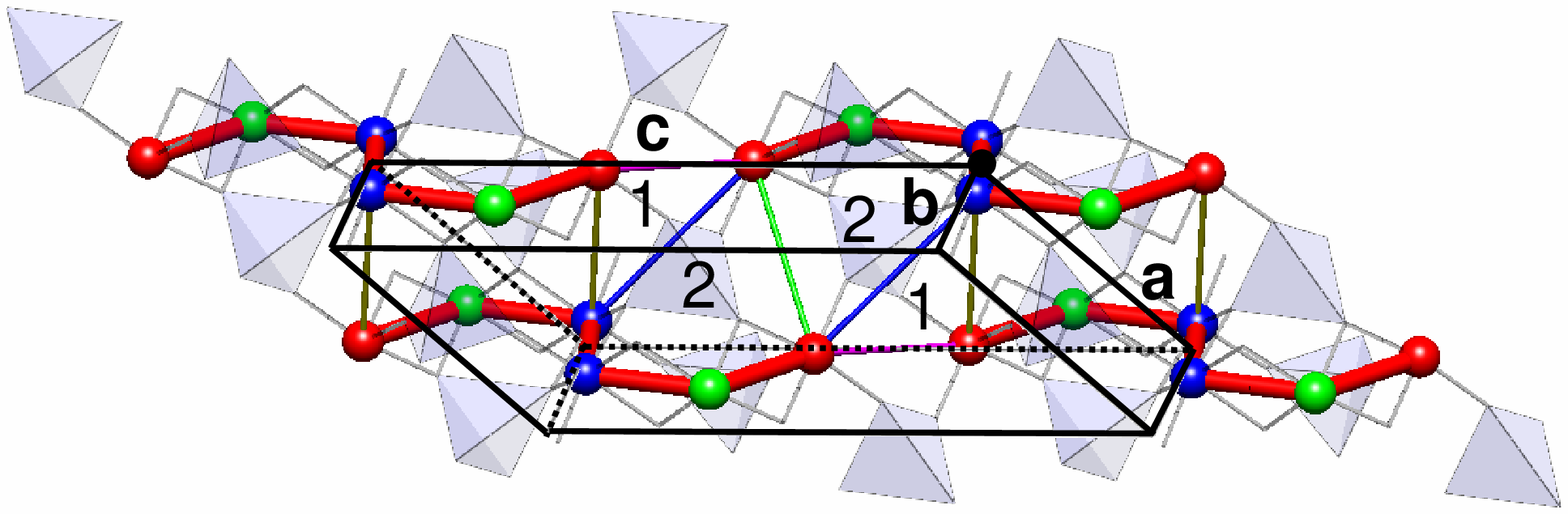}
\caption{(Color) (a) Crystal structure and magnetic exchange frustration in \fevo. The basic magnetic unit are \emph{S}-shaped clusters made of 6 \fe \ atoms (red, blue, and green large balls) well spaced by V$^{5+}$O$_4$ tetrahedra (blue). The Fe-O bonds are drawn as thin grey lines (O atoms are not shown for clarity). The thick red lines represent intra-cluster interactions, with contributions from direct Fe-Fe exchange and Fe-O-Fe super-exchange. Iron atoms belonging to different clusters are coupled by Fe-O-O-Fe SSE paths mediated by the edges of the VO$_4^{3-}$ tetrahedra giving rise to effective inter-cluster Fe-Fe interactions (thin colored lines)}
\label{fig1}
\end{figure}
\indent The crystal structure of \fevo \ is shown in Fig. \ref{fig1}.  To facilitate the description of the magnetic structures (see below) we have redefined the triclinic basis vector \textbf{c} so that $\textbf{c}=-2\textbf{a'}-\textbf{c'}$, where \textbf{a'} and \textbf{c'} are the basis vectors used in Ref.\onlinecite{Robertson1972}. The important crystallographic features  for the description of the magnetic properties are easily identified. The Fe$^{3+}$ ions, all in the high-spin state S=5/2 (see below), are arranged in clusters, separated by (VO$_4$)$^{3-}$ groups, containing non-magnetic V$^{ 5+}$ ions.  Each cluster of 6 Fe$^{3+}$ ions consists of two identical Fe$_3$O$_{13}$ monomers, related by a center of inversion.  With our convention, the \textbf{c}- axis runs along the line connecting adjacent clusters through their centers and open ends.  In Fig\ref{fig1}, we have also represented the two main types of magnetic interactions: intra-cluster direct-exchange and SE interactions (thick lines), which are antiferromagnetic (AF), and inter-cluster SSE interactions (thin lines). There is clearly the potential for magnetic frustration, since Fig.\ref{fig1} reveals that the effective inter-cluster and intra-cluster Fe-Fe linkages  form loops (label 1 and 2), which contain an odd number of Fe sites, within which collinear AF arrangements cannot be fully satified.\\
\indent The magnetic frustration and competition between SE and SSE interactions give rise to a complex low temperature phase diagram, reminiscent of that of other cycloidal magnets. Above 100~K the magnetic susceptibility follows a Curie-Weiss law (not shown). The effective paramagnetic moment $\mu_{eff}=6.103(4)\mu_{B}$ is near the value expected for Fe$^{3+}$ ions in the high-spin S=5/2 state ($\mu^{th}_{eff}=5.91\mu_{B}$).  The Weiss temperature $\theta_{CW}=-124.9(4)K$ confirms the presence of strong AF interactions. The low temperature data clearly indicate  magnetic transitions at T$_{N1}$=22K and T$_{N2}$=15K, consistent with a frustration index of $|\theta_{CW}|/T_{N1}\sim 6$. The specific heat, (Fig. \ref{fig2} a) also shows two lambda-type anomalies at these critical temperatures. The magnetic contribution to the specific heat was determined by subtracting the lattice contribution, indicated by the dashed line in Fig. \ref{fig2} a), which was estimated by fitting the data at high temperature to the Debye function (Debye temperature of 385(5)K)). The magnetic entropy integrated between 2K and 50K is $S_M=$13.98 J.K$^{-1}$.mol$^{-1}$, approaching the classical value of $R\log(2S+1)=$14.89 J.K$^{-1}$.mol$^{-1}$  for S=5/2.  A large fraction of the total magnetic entropy ($\sim30\%$) is only recovered far above T$_{N1}$, indicative of short-range AF ordering above \tna\cite{Comment}. \\
\begin{figure}[b]
\includegraphics[height=240pt,angle=-90]{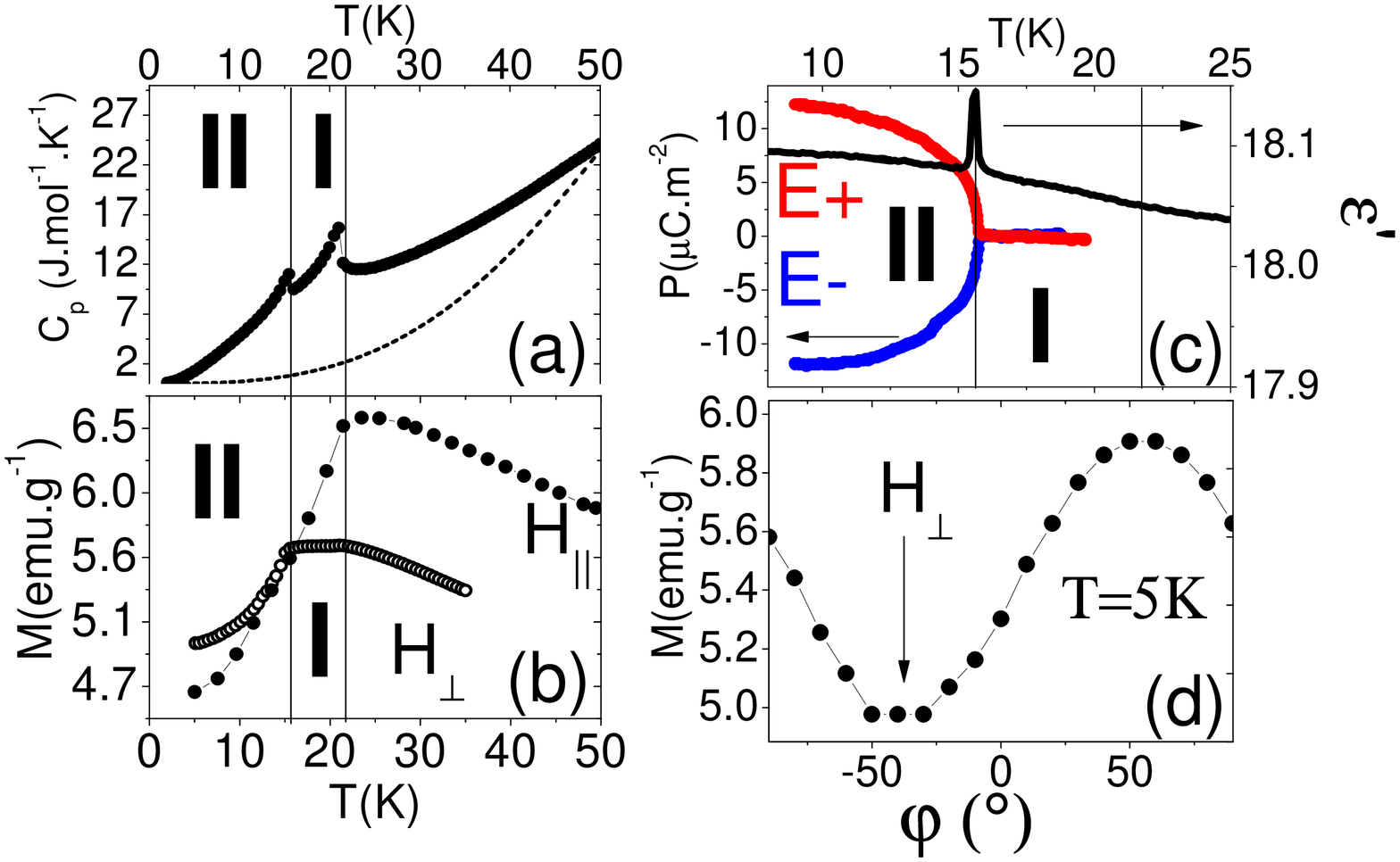} \\
\caption{Magneto-electric phase diagram for FeVO$_4$. (a) Heat capacity (filled circles). The dashed line shows the estimated lattice Debye contribution (see text for details) (b) Single crystal magnetization at 3 Tesla with the magnetic field applied in the directions \textbf{H}$_{\parallel}$(filled circles) and \textbf{H}$_{\perp}$ (opened circles) explicated in the panel (e). c) Bulk electric polarization measured for a sample cooled in a positive (E+,red) and negative (E-,blue) electric field of 160V and dielectric constant (black solid line) d) Angular dependence of the magnetization at 3 Tesla and T=5K measured varying the orientation of the applied magnetic field  in the $(\textbf{b}^{*},\textbf{c}^{*})$ plane.  $\mathbf{H}$ is aligned with $\mathbf{c}^{*}$ when $\phi=0$.} \label{fig2}
\end{figure}
\indent To obtain further insight into the nature of the magnetic ordering taking place at T$_{N1}$ and T$_{N2}$ ,  magnetic measurements were made on the needle-shaped single crystals. The crystal orientation was set at 300K before cooling the crystal in a 3-T field and recording magnetization data between 5 and 35K. At low temperatures, the direction of minimal susceptibility coincides with the direction of growth of the needle-shaped crystals (approximately along the crystallographic direction $\mathbf{a}$). With this field orientation, labeled \textbf{H}$_{\parallel}$ in Fig. \ref{fig2}b, one only observes the magnetic transition at T$_{N1}$. The transition at T$_{N2}$ is clearly visible in measurements with the field applied \emph{perpendicular} to the $\mathbf{a}$ axis, i.e., in the $\mathbf{b}^* - \mathbf{c}^*$ plane.  By rotating the crystal around its needle axis, we have determined the secondary easy magnetic directions in this plane.  This is shown in Fig. \ref{fig2}d.   The minimum and maximum of the magnetization were found for the crystal rotated $\phi_{min}=-40\grad$ and $\phi_{max}=50\grad$ away from the $\mathbf{c}^*$ direction, respectively. The temperature dependance of the magnetization with the field orientation $\phi_{min}$, labeled as \textbf{H}$_{\perp}$ in Fig. \ref{fig2}b shows a pronounced drop  below T$_{N2}$\cite{Comment1}. The phase boundaries indicated in Fig. \ref{fig2} therefore mark the domains of stability of two distinct magnetic phases (I and II).  The temperature dependence of the dielectric susceptibility, shown in Fig. \ref{fig2}(c), has no anomaly at T$_{N1}$ but the sharp peak at T$_{N2}$ and the concomitant appearance of an electrical polarization (Fig. 2c) demonstrates that the system becomes ferroelectric in phase II, while phase I is not ferroelectric. The value of the electrical polarization at low temperatures is an order of magnitude smaller than for TbMnO$_3$\cite{Kimura2003}, a large reduction which can be partly attributed to the fact that our dielectric measurements were made on polycrystalline samples. \\
\begin{figure}[tpb]
\includegraphics[width=240pt]{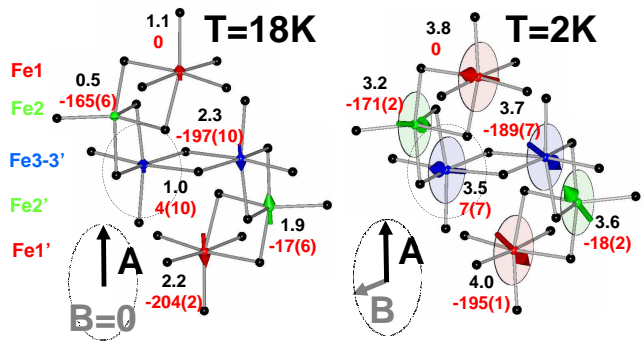}
\caption{ (Color) Zoom in on the magnetic order of the Fe \emph{S-}shaped cluster of Fig.\ref{fig4} containing the surrounded spin.  The three inequivalent Fe sites are shown as small red, green and blue spheres, and the Fe sites with primed  labels are obtained by inversion symmetry. The Fe are arranged in two types of edge-sharing polyhedra: two \fe O$_{5}$ trigonal bypiramids (sites 2-2') and the 4 other sites in \fe O$_{6}$ octahedra. Spin are aligned with a common direction $\mathbf{A}$ at T=18K, and rotate in the same (\textbf{A,B}) plane at T=2K. The magnetic moment amplitudes (black labels) were calculated using equation \ref{eq1} with the refined parameters given in the supplementary information\cite{Suppl}, accounting for a de-phasing term of $68.4\grad$ coming from the position of the cluster in the crystal ($\mathbf{R}_n=(0 \ \bar{2} \ 1)$). Only the values of the refined phases $\varphi_j$ are given in degrees (red labels) for clarity.}
\label{fig3}
\end{figure}
\begin{figure}[tpb]
\includegraphics[width=240pt]{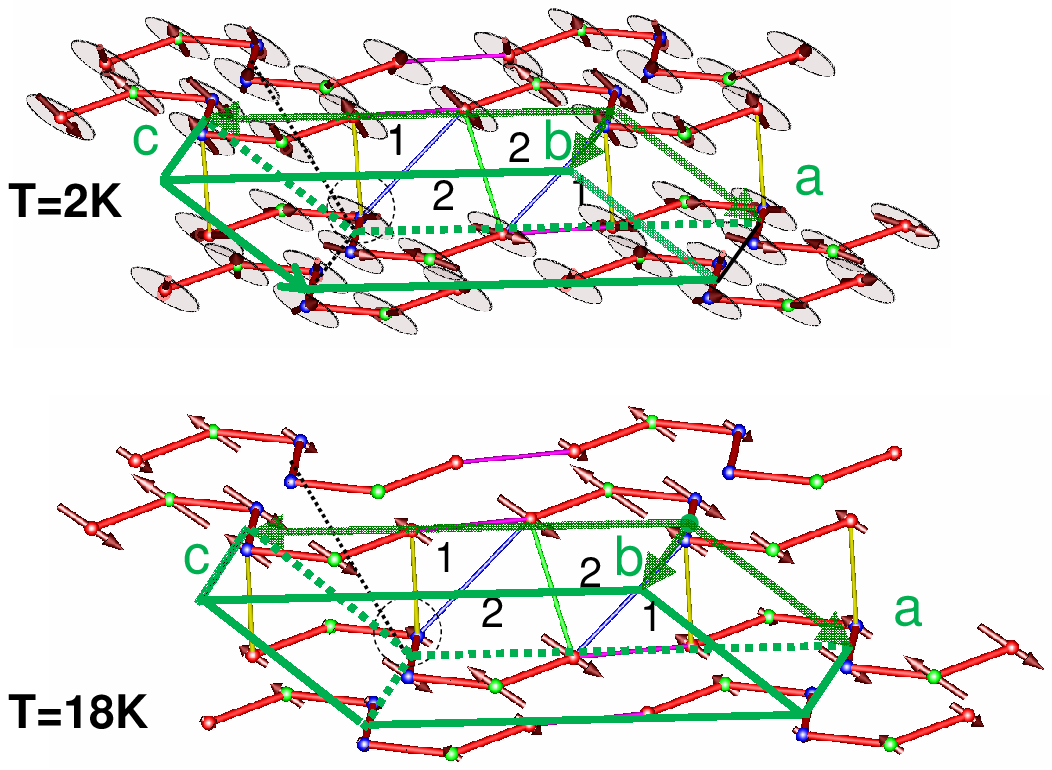}
\caption{ (Color) Magnetic structure of \fevo \ as derived from refinements of the single crystal neutron diffraction data $T=2K$ (top) and $T=18K$ (bottom).  The magnetic moments are shown as brown arrows. For the non-collinear structure at 2K, the envelop of the helicoidal modulation is also shown by grey ellipses at each lattice site. Frustrated exchange loops are reproduced form Fig.\ref{fig1}, and one spin on loop 1 experiencing the geometrical exchange frustration is surrounded.}
\label{fig4}
\end{figure}
\indent The magnetic structures of phases I and II were determined by single-crystal neutron diffraction. Below T$_{N1}$, the data show new Bragg peaks of magnetic origin, which can all be indexed with a single, nearly temperature-independent propagation vector $\mathbf{k}=(0.222 \ -0.089\ 0.012)$ almost perpendicular to $\mathbf{c}$.  In the most general case of a single-$\mathbf{k}$ incommensurately modulated structure, the magnetic moments \textbf{M$_j$(R$_l$)} on a given crystallographic site describe an ellipse as they propagate in different unit cells \textbf{R$_l$}. \textbf{M$_j$(R$_l$)} can therefore be written as:
\begin{equation}
\bf{M}_{j}(R_l) = {\bf{A}_{j}} \cos({2\pi.\bf{k\cdot R_{l}}}+\varphi_{j})+{\bf{B}_{j}} \sin({2\pi.\bf{k\cdot R_{l}}}+\varphi_{j})
\label{eq1}
\end{equation}
where \textbf{A$_j$} and  \textbf{B$_j$} are two perpendicular vectors defining the major and minor semiaxes of the ellipse\cite{Zaharko2004}.  An initial set of refinements  indicated that the magnetic structure of phase I is a \emph{collinear} spin-density wave (i.e., \textbf{B$_j$}$=0$), whereas phase II possesses a \emph{helical} structure.  It is noteworthy that the collinear direction of the spins in phase I coincides with the major semiaxis \textbf{A$_j$} of the ellipses in phase II, indicating that the spin anisotropy does not change at the phase boundary.  In principle, both the amplitudes and directions of \textbf{A$_j$} and  \textbf{B$_j$} as well as the phase angles of the modulations can be different for each site.  We have shown that the  correlation between  parameters can be significantly reduced by introducing constraints \cite{Suppl} whilst still yielding refinements of excellent quality. This has lead us to consider only minimal models for phase I (10 parameters, $R_{F2}=10.3\%$ and $\chi^2=1.7$) and phase II (14 parameters, $R_{F2}=5.07\%$ and  and $\chi^2=8.2$). The constraints retained suggest that the data do not support significantly different orientations of the ellipses and that the amplitudes $\mid$\textbf{A$_j$}$\mid$ and $\mid$\textbf{B$_j$}$\mid$ (fig.\ref{fig3}) of atoms related by centro-symmetry do not differ significantly, although their phases are not related by centro-symmetry and need to be determined independently. Perspective views of the magnetic structures of phases I and II are shown for one cluster in Fig. \ref{fig3} and for few clusters in Fig.\ref{fig4}. \\
\indent The helical magnetic order reveals different degree of frustration in different directions. It is characterized by the presence of  quasi-1D AF order on chains of S-shaped clusters running in the $\mathbf{c}$ direction. The SSE path linking  clusters in this direction is therefore probably the least frustrated SSE interaction inducing only very slow rotations of the average AF direction over a very long period of approximately 110 nm. This contrasts with all the other SSE pathes, which induce large rotations of the average AF direction between neighboring chains. \\
\indent Noteworthily, none of the magnetic orders is centrosymmetric because the difference between the phases of sites related by inversion symmetry deviates significantly from 180$\grad$ and similar phases are refined at T=2K and T=18K (Fig.\ref{fig3}). This inversion symmetry breaking is clearly visible at T=2K. Each half cluster of 3 Fe shows an almost perfect AF order, but between the two halves, the AF directions are canted by $15\sim20\grad$. The collinear magnetic order at T=18K simply appears as the projection of the helical order on the common \textbf{A} direction. We argue that the observed \textbf{k} vector and the refined phases are here fixed by the symmetric part of the Heisenberg magnetic exchange energy independently of the presence or not of a \textbf{B} component, and that other energy terms control the stability of the collinear state with respect to the non collinear state. \\
\indent The transition to collinear magnetic order in multiferroic magnets with a spiral ground state has already been discussed by Mostovoy\cite{Mostovoy2006}. At the temperature \tnb, there is a cross-over  between a high temperature regime in which anisotropy dominates so that the moments align along the local easy axis, and one in which competition between anisotropy and entropy favors disordered components of moment in directions orthogonal to it. For \fevo \ the local easy axis is the \textbf{A} direction deduced from the neutron data, which coincides with the macroscopic easy axis,  labelled \textbf{H}$_\parallel$ determined from the magnetisation measurements and labelled \textbf{H}$_\parallel$ in Fig. 2b. In phase II the moments are confined to the plane of the ellipses with \textbf{A} and \textbf{B} as semi-axes, drawn in Figs 3 and 4. When a field is applied in any direction orthogonal to \textbf{A} (\textbf{H}$_\parallel$) the  temperature dependence of the magnetisation shows a singularity at \tnb, a rapid fall below \tnb\  and almost no singularity at \tna. The orthogonal direction for which the fall is most pronounced is parallel to \textbf{B}, the minor axis of the ellipse. The non-collinearity of phase II is due to the stabilization of the second order parameter modulating moments  oriented parallel to \textbf{B}; it reflects the presence of additional  terms in the free energy which outweigh the entropy effect. The non-collinearity suggests that these terms are due to antisymmetric spin-orbit coupling  of the Dzyaloshinski-Morya type proportional to vector products of spins $\mathbf{S}_i\times \mathbf{S}_j$. Most theories of multiferroic spiral magnets have invoked this interaction as the driving force leading to magnetically induced ferro-electricity. However it is not yet clear whether it is spin currents associated with these terms\cite{Katsura2005}, or the simultaneous  emergence of ion displacements linked to configurational asymmetry in the magnetic structure,\cite{Xiang2008} which give the larger contribution to the ferroelectric polarisation.  \\
\indent In summary, we have demonstrated magnetically induced ferroelectricity in \fevo. It is noteworthy that its magnetoelectric phase diagram in the vicinity of \tnb \ is very similar to that of the classical TbMnO$_3$ system in the vicinity of T$_{lock}$=27K\cite{Kimura2003,Prokhnenko2007a}. This gives a strong indication that  spin-orbit interactions may make an important contribution to the electrical polarization in \fevo. However, the $Fe^{3+}$ ion has  L=0 in the free-ion state and in \fevo\ should be orbitally quenched which reinforces the \emph{induced} nature of the mechanisms responsible for multiferroic behavior in spiral magnets. The magneto-electric coupling is simpler in \fevo, because the only magnetic atoms
participating are Fe; it therefore  not influenced or complicated by rare earth magnetism, as this is the case in TbMnO$_3$\cite{Prokhnenko2007a}. \\
\indent This study of \fevo\ has clarified the relative roles of exchange frustration, anisotropy, entropy, and antisymmetric exchange in defining the magnetic state of ``type-II'' multiferroics.  In TbMnO$_3$ it is difficult to identify the origin of the magnetic frustration because the SSE interactions form a network of next nearest neighbor interactions interpenetrating that of the first neighbor superexchange interactions. In \fevo\ on the other we have established  a clear connection between the moment reduction or the spin rotations, which are typical of  the incommensurate magnetic phases of spiral magnets, and the most relevant SSE paths giving rise to frustration.

\indent We would like to thank Dr J. A. Rodriguez-Velazman for his support during the neutron diffraction experiment.

\bibliographystyle{apsrev}

\begin{thebibliography}{20}
\bibitem[{Com({\natexlab{a}})}]{Comment}
\bibinfo{note}{This observation is consistent with a strong magnetic diffuse
  scattering signal above \tnb, in neutron powder diffraction experiments,
  which we also performed separately in the T$_{N1}<$T$<$30K temperature range.
  A more detailed account of all these observations will be reported elswhere}.
\end{thebibliography}

\end{document}


\title{Supplementary information: Multiferroicity and spiral magnetism in FeVO$_4$ with quenched Fe orbital moments}
\date{\today}


\maketitle

\indent In the case of \fevo, the modulation and the symmetry of magnetic structures is more conveniently described in an alternative cell setting $(\mathbf{a},\mathbf{b},\mathbf{c})=(\mathbf{a'},\mathbf{b'},\mathbf{-2a'-c'})$, where $(\mathbf{a'},\mathbf{b'},\mathbf{c'})$ was the cell used in the original structural description given in Ref.\onlinecite{Robertson1972}. In the original work,  cell parameter at RT are $\mathbf{a'}=6.71633 \ \AA$, $\mathbf{b'}=8.06817 \ \AA$, $\mathbf{c'}=9.34715 \ \AA$, $\alpha'=96.6537\grad$,$\beta'=106.528\grad$, $\gamma'=101.499\grad$, so that with our new setting, $\mathbf{a}=6.71633 \ \AA$,     $\mathbf{b}=8.06817 \ \AA$, $\mathbf{c}=14.01312 \ \AA$, $\alpha=74.432\grad$,     $\beta=140.248\grad$, $\gamma=101.499\grad$.
The transformed Fe atomic positions in this new cell are given in table \ref{pos}. \\
\begin{table}[h]
\begin{center}
\begin{tabular}[b]{*{4}{c}}
\hline \hline
Atom  &   x  &  y  &  z  \\
\hline
  Fe$_1$  & -0.06433  &  0.69426  & -0.40839\\
  Fe$_2$  &  0.04169  &  0.89000  & -0.21254\\
  Fe$_3$  &  0.05455  &  0.69432  &  0.01143\\
  Fe$_4$  & -0.05450  &  0.30573  & -0.01143\\
  Fe$_5$  & -0.04169  &  0.11000  &  0.21250\\
  Fe$_6$  &  0.06443  &  0.30584  &  0.40839\\
\end{tabular}
\end{center}
\caption{Atomic positions for \fevo \ in the transformed cell $(\mathbf{a},\mathbf{b},\mathbf{c})$ (see text)}
\label{pos}
\end{table} \\
\indent The magnetic moments \textbf{M$_j$(R$_l$)} of any site j in a unit-cell translated by a vector $R_l$ with respect to the origin is written:
\begin{equation}
\bf{M}_{j}(R_l) = {\bf{A}_{j}} \cos({2.\pi\bf{k\cdot R_{l}}}+\varphi_{j})+{\bf{B}_{j}} \sin({2.\pi\bf{k\cdot R_{l}}}+\varphi_{j})
\label{eq1}
\end{equation}
where the constraint (\textbf{A$_j$}$\perp$\textbf{B$_j$}) is always imposed in our refinements. In \fevo, the propagation vector is $\mathbf{k}=(0.222 \ -0.089\ 0.012)$. \\
\indent In the program Fullprof\cite{Rodriguez-Carvajal1993}, we can constrain the direction of the parameters describing the magnetic structure. This is best done using spherical coordinates attached to a $(x,y,z)$ orthogonal frame and defined in a way such as x coincides with $\mathbf{a}$ and z coincides with $\mathbf{c^{*}}$. \\
\indent For amplitude modulated phases (\textbf{B$_j$}=0) (table \ref{table2}), we use the conventional latitude and azimuth spherical angles $(\phi_j,\theta_j)$ with $(\phi_j,\theta_j)=(0,0)$ giving z, and $(\phi_j,\theta_j)=(0,90\grad)$ giving x, in order to parameterize and/or constrain the \textbf{A$_j$} directions . \\
\indent For helicoidal structures, we used the option in Fullprof which orients the vectors (\textbf{A$_j$},\textbf{B$_j$}) and the vector \textbf{N$_j$} normal to the (\textbf{A$_j$},\textbf{B$_j$}) plane along the axis of orthogonal coordinate systems $(x',y',z')$ obtained by applying active rotations given by the Euler angles $(\phi_j,\theta_j,\chi_j)$, which rotates the $(x,y,z)$ coordinate frame in turn along z, y, and z again. The parameters $\chi_j$ are in fact always constrained to be the same for all sites to avoid correlations with the phases $\varphi_{j}$ (table \ref{table3}).
\begin{table*}[h]
\begin{center}
\begin{tabular}[b]{*{5}{c}}
\hline
\hline
\multicolumn{5}{c}{Model I: fixed directions  $\textbf{A}$ } \\
\multicolumn{5}{c}{ for all sites and symmetry constraints on phases} \\
\multicolumn{5}{c}{($R_{B}=38.9\%$ $\chi^2=32.4$ : not converging) } \\
\hline
 Atom & $|\textbf{A}|$  & $\phi(\grad)$ & $\theta(\grad)$ & $\varphi(\grad)$ \\
\hline
Fe$_1$   & 3.5(2)   &  $\phi_1=6(3)$   &    $\theta_1=94(3)$    & 0             \\
Fe$_2$   & 2.9(2)   &  $\phi_2=\phi_1$ &    $\theta_2=\theta_1$ & 167(2)        \\
Fe$_3$   & 2.8(2)   &  $\phi_3=\phi_1$ &    $\theta_3=\theta_1$ & -20(3)        \\
Fe$_4$   & $|A_3|$  &  $\phi_4=\phi_1$ &    $\theta_4=\theta_1$ & $\varphi_3+180$ \\
Fe$_5$   & $|A_2|$  &  $\phi_5=\phi_1$ &    $\theta_5=\theta_1$ & $\varphi_2+180$ \\
Fe$_6$   & $|A_1|$  &  $\phi_6=\phi_1$ &    $\theta_6=\theta_1$ & $\varphi_1+180$ \\
\hline
\hline
\multicolumn{5}{c}{Model II: model I + } \\
\multicolumn{5}{c}{ patial relaxation of the spin directions} \\
\multicolumn{5}{c}{($R_{B}=39.9\%$ $\chi^2=33.7$: not converging) } \\
\hline
 Atom & $|\textbf{A}|$  & $\phi(\grad)$ & $\theta(\grad)$ & $\varphi(\grad)$ \\
\hline
Fe$_1$   & 3.5(2) & 10(6)                 &  87(7)                &  0      \\
Fe$_2$   & 2.7(3) & 4(7)                  & 98(8)                 & 169(2)  \\
Fe$_3$   & 2.7(2) & 5(7)                  & 96(9)                 & -20(3)  \\
Fe$_4$   & $|A_3|$ & $\phi_4=\phi_3$ & $\theta_4=\theta_3$  &$\varphi_3+180$\\
Fe$_5$   & $|A_2|$ & $\phi_5=\phi_2$ & $\theta_5=\theta_2$  &$\varphi_2+180$\\
Fe$_6$   & $|A_1|$ & $\phi_6=\phi_1$ & $\theta_6=\theta_1$  &$\varphi_1+180$ \\
\hline
\hline
\multicolumn{5}{c}{Model III: Model I + }\\
\multicolumn{5}{c}{relaxation of the symmetry constraints on phases} \\
\multicolumn{5}{c}{($R_{B}=  10.3     \%$ $\chi^2= 1.64$) } \\
\hline
 Atom & $|\textbf{A}|$  & $\phi(\grad)$ & $\theta(\grad)$ & $\varphi(\grad)$ \\
\hline
Fe$_1$   & 3.01(5) &  $\phi_1=3.9(5)$   &   $\theta_1=95.6(6)$  & 0          \\
Fe$_2$   & 3.23(5) &  $\phi_2=\phi_1$   &   $\theta_2=\theta_1$ & -165(6)    \\
Fe$_3$   & 2.86(3) &  $\phi_3=\phi_1$   &   $\theta_3=\theta_1$ & 4(10)      \\
Fe$_4$   & $|A_3|$ &  $\phi_4=\phi_1$   &   $\theta_4=\theta_1$ & -197(10)   \\
Fe$_5$   & $|A_2|$ &  $\phi_5=\phi_1$   &   $\theta_5=\theta_1$ & -17(6)     \\
Fe$_6$   & $|A_1|$ &  $\phi_6=\phi_1$   &   $\theta_6=\theta_1$ & -204(2)    \\
\hline
\hline
\multicolumn{5}{c}{Model IV: model I + } \\
\multicolumn{5}{c}{relaxation of the spin directions (partial)} \\
\multicolumn{5}{c}{ and phases symmetry constraints} \\
\multicolumn{5}{c}{($R_{B}=10.2  \%$ $\chi^2= 1.62 $)  } \\
\hline
 Atom & $|\textbf{A}|$  & $\phi(\grad)$ & $\theta(\grad)$ & $\varphi(\grad)$ \\
\hline
Fe$_1$  &  3.00(5) & 3(2) & 95(2)                              & 0        \\
Fe$_2$  &  3.23(5) & 7(2) & 94(2)                              & -168(9)  \\
Fe$_3$  &  2.88(4) & 1(2) & 97(2)                              & 3(2)     \\
Fe$_4$  & $|A_3|$ & $\phi_4=\phi_3$ & $\theta_4=\theta_3$      & -197(2)  \\
Fe$_5$  & $|A_2|$ & $\phi_5=\phi_2$ & $\theta_5=\theta_2$      & -20(9)   \\
Fe$_6$  & $|A_1|$ & $\phi_6=\phi_1$ & $\theta_6=\theta_1$      & -203(2)   \\
\end{tabular}
\end{center}
\caption{Magnetic parameters refined at T=18K}
\label{table2}
\end{table*}

\begin{table}[h!]
\begin{center}
\begin{tabular}[b]{*{7}{c}}
\hline
\hline
\multicolumn{7}{c}{Model I: fixed directions for all sites and symmetry constraints on phases} \\
\multicolumn{7}{c}{($R_{B}=27.9\%$ $\chi^2=270.0$: not converging) } \\
\hline
  Atom      &   $|A|$  &  $|B|$  &  $\chi(\grad)$  &$\phi(\grad)$    &
                  $\theta(\grad)$ & $\varphi(\grad)$  \\
\hline
FE$_1$ & 4.0(5) & 4.6(5) &   $\chi_1= -55(5)$  &   $\phi_1=87(6)$  &  $\theta_1=41.3(6)$   & 0       \\
FE$_2$ & 3.1(5) & 3.3(5) &   $\chi_2=\chi_1$   &   $\phi_2=\phi_1$ &  $\theta_2=\theta_1$  & 188(3)   \\
FE$_3$ & 5.0(4) & 3.3(5) &   $\chi_3=\chi_1$  &   $\phi_3=\phi_1$ &  $\theta_3=\theta_1$  &  21(2)   \\
FE$_4$ & $|A_3|$ & $|B_3|$ & $\chi_4=\chi_1$  &   $\phi_4=\phi_1$ &  $\theta_4=\theta_1$  & $\varphi_3+180$ \\
FE$_5$ & $|A_2|$ & $|B_2|$ & $\chi_5=\chi_1$  &   $\phi_5=\phi_1$ &  $\theta_5=\theta_1$  & $\varphi_2+180$ \\
FE$_6$ & $|A_1|$ & $|B_1|$ & $\chi_6=\chi_1$  &   $\phi_6=\phi_1$ &  $\theta_6=\theta_1$  & $\varphi_1+180$ \\
\hline
\hline
\multicolumn{7}{c}{Model II: model I + relaxation of the phases symmetry constraints} \\
\multicolumn{7}{c}{($R_{B}= 5.07\%$ $\chi^2=8.24$) } \\
\hline
  Atom      &   $|A|$  &  $|B|$  &  $\chi(\grad)$  &$\phi(\grad)$    &
                  $\theta(\grad)$ & $\varphi(\grad)$  \\
\hline
FE$_1$ & 4.51(7) & 3.67(8)  &  $\chi_1=-77(2)$  & $\phi_1=81(2)$   &  $\theta_1=33.2(1)$    &  0         \\
FE$_2$ & 4.18(6) & 3.16(7)  &  $\chi_2=\chi_1$  & $\phi_2=\phi_1$  &  $\theta_2=\theta_1$   &  -171(2)    \\
FE$_3$ & 4.29(7) & 3.39(9)  &  $\chi_3=\chi_1$  & $\phi_3=\phi_1$  &  $\theta_3=\theta_1$   &  7(7)      \\
FE$_4$ & $|A_3|$ & $|B_3|$  &  $\chi_4=\chi_1$  & $\phi_4=\phi_1$  &  $\theta_4=\theta_1$   &  -189(7)   \\
FE$_5$ & $|A_2|$ & $|B_2|$  &  $\chi_5=\chi_1$  & $\phi_5=\phi_1$  &  $\theta_5=\theta_1$   &  -18(2)    \\
FE$_6$ & $|A_1|$ & $|B_1|$  &  $\chi_6=\chi_1$  & $\phi_6=\phi_1$  &  $\theta_6=\theta_1$   &  -195.2(7)  \\
\hline
\hline
\multicolumn{7}{c}{Model III: Model II + partial relaxation of the directions constraints} \\
\multicolumn{7}{c}{($R_{B}= 6.29  \%$ $\chi^2=12.7$ : slightly diverging) } \\
\hline
  Atom      &   $|A|$  &  $|B|$  &  $\chi(\grad)$  &$\phi(\grad)$    &
                  $\theta(\grad)$ & $\varphi(\grad)$  \\
\hline
FE$_1$ & 4.52(8) &  3.7(1)  &  $\chi_1= -76(2)$   &  76(2) & 20.7(5)                             &   0          \\
FE$_2$ & 4.21(8) & 3.06(9)  &  $\chi_2=\chi_1$    &  73(2) & 28.8(3)                             &   -167(9)    \\
FE$_3$ & 4.31(8) &  3.4(2)  &  $\chi_3=\chi_1$    &  85(3) & 42.6(3)                             &   0(7)       \\
FE$_4$ & $|A_3|$ & $|B_3|$  &   $\chi_4=\chi_1$   & $\phi_4=\phi_3$ & $\theta_4=\theta_3$        &  -196(7)     \\
FE$_5$ & $|A_2|$ & $|B_2|$  &   $\chi_5=\chi_1$   & $\phi_5=\phi_2$ & $\theta_5=\theta_2$        &  -15(9)      \\
FE$_6$ & $|A_1|$ & $|B_1|$  &   $\chi_6=\chi_1$   & $\phi_6=\phi_1$ & $\theta_6=\theta_1$        &  -196.2(9)   \\
\end{tabular}
\end{center}
\caption{Magnetic parameters refined at T=2K}
\label{table3}
\end{table}

\bibliographystyle{apsrev}